\documentclass[twocolumn,usenatbib,amsmath,amssymb,amsbsy]{mn2e}

\usepackage{natbib}
\usepackage{amsbsy}
\usepackage{amssymb}
\usepackage{amsmath}

\usepackage{graphicx}

\newcommand{\be}{\begin{equation}}
\newcommand{\ee}{\end{equation}}
\newcommand{\bear}{\begin{eqnarray}}
\newcommand{\eear}{\end{eqnarray}}
\newcommand{\n}{\hat{n}}

\newcommand{\m}{\hat{m}}

\newcommand{\rc}{{\rm c}}
\newcommand{\rs}{{\rm s}}

\newcommand{\rC}{{\rm c}}
\newcommand{\rB}{{\rm B}}

\newcommand{\eB}{{\epsilon_{\rm B}}}
\newcommand{\ed}{{\epsilon_{\rm c}}}
\newcommand{\eA}{{\epsilon_{\rm A}}}

\newcommand{\eS}{{\epsilon_{\rm s}}}
\newcommand{\nB}{{\n_{\rm B}}}
\newcommand{\nOm}{{\n_{\rm \Omega}}}
\newcommand{\nd}{{\n_{\rm c}}}

\newcommand{\nS}{{\n_{\rm s}}}

\begin{document}

\title[Magnetar non-precession] {Implications of magnetar non-precession}

\author[Glampedakis, \& Jones]{K. Glampedakis$^1$ \& D.I. Jones$^2$ \\
  \\
  $^1$  Theoretical Astrophysics, University of Tuebingen, Auf der Morgenstelle 10, Tuebingen, D-72076, Germany \\
  $^2$ School of Mathematics, University of Southampton, Southampton
  SO17 1BJ, UK}

\maketitle

\begin{abstract}

The objects known as anomalous X-ray pulsars and soft gamma repeaters are commonly identified with magnetars, neutron
stars with ultrastrong magnetic fields. The rotational history of these objects has, so far, revealed no evidence of
free precession. At the same time  these objects do not generally appear to have magnetic axes nearly parallel or orthogonal 
to their spin axes. In this paper we show that the combination of these two observations, together with simple rigid-body dynamics, 
leads to non-trivial predictions about the interior properties of magnetars: either (i) elastic stresses in magnetar crusts are close 
to the theoretical upper limit above which the crustal matter yields or (ii) there is a ``pinned'' superfluid component 
in the magnetar interior.  As a potentially observable consequence of these ideas we  point out that, in the case of no pinned 
superfluidity, magnetars of stronger magnetic field strength than those currently observed would have to be nearly 
aligned/orthogonal rotators.

\end{abstract}


%%%%%%%%%%%%%%%%%%%%%%%%%%%%%%%%%%%%%%%%%%%%%
\section{Introduction}
\label{sec:intro}

The magnetars are a subset of the neutron star population with ultrastrong magnetic fields, typically greater than $10^{14}$ G, 
and long spin periods, typically $1-10$ seconds.  They manifest themselves observationally as anomalous X-ray pulsars (AXPs) 
and soft gamma repeaters (SGRs); see \citet{woods} for a review.  
The latter class of objects are of particular interest, as periodicities have been observed in the 
tails of bursts from some objects, possibly providing the first ever evidence for excitation of neutron star normal modes.  
The theory of magnetars as very strongly magnetised neutron stars was developed by Thompson \& Duncan (see e.g. \citet{TD}), in which the 
decaying magnetic field powered the  outbursts.

In this paper we will concern ourselves with something that the magnetars \emph{don't do}---they do not seem to undergo free precession.  
More precisely, we will base our analysis on the following two working assumptions, both motivated directly by observations:  
(i) magnetars do not precess,  and (ii) their magnetic pulsation axes are neither aligned nor orthogonal to the spin axes.   
On the basis of these two observational results and simple rigid body dynamics we can then deduce that 
\emph{either} (a) the stellar crust is very highly strained, right at or close to the limit predicted theoretically 
by detailed modelling, \emph{or} (b) the magnetars contain a pinned superfluid component. We believe that either conclusion is interesting.

The first assumption, the lack of precession, follows from timing studies.  
Precession would result in a periodic modulation in the time of arrival (and possibly also shape) of the magnetar  pulsations \citep{ja01}.  
Indeed, early observations of timing irregularities prompted Melatos to suggest precession was a generic feature of the 
AXP population \citep{melatos}.  However, further observation of the AXPs \citep{kaspi1, kaspi2, gavriil} and 
SGRs \citep{woods_etal} has ruled out precession at an amplitude level above the rms amplitude of 
magnetar timing noise. Small-amplitude precession, `buried' inside the timing noise, is still a possibility and future timing 
data analysis should attempt to address this issue. (A strict lack of precession is not essential for our arguments: 
as will become clear, a small precession amplitude would imply that the magnetar spins about an axis close to a principal axis or 
the superfluid pinning axis; this is the essential point for our modelling).

Our second assumption concerns the magnetar inclination angles, i.e. the angle $\theta_B$ between their spin axes and magnetic axes; 
we assume that magnetars are neither aligned ($\theta_B \approx 0$) nor orthogonal ($\theta_B \approx \pi/2$).  
By talking of a `magnetic axis', we are implicitly assuming there exists a well defined axis about which the internal magnetic field 
is approximately symmetric; numerical simulations indicate this is likely to be the case \citep{braith09}.  
Then, to make any observationally-motivated statement about the location of this axis, we also assume that this axis coincides 
with the axis defined by the magnetar pulsations, which is probably a reasonable assumption given the known importance of the 
magnetic field in determining the surface temperature profile \citep{opk01}.  
 
Given these assumptions,  the result that the magnetars are not aligned rotators follows trivially from the fact  that we see 
magnetar pulsations at all.   That the two axes are not orthogonal can be argued on the basis of the harmonic content of the pulse profiles.
An orthogonal rotator with two nearly identical anti-podal hot spots would, regardless of the observer's location, display a strong second 
harmonic in its pulse power spectrum, something which has not been reported to be generic (but see the data for 1E 2259.1+586 in Fig. 2 
of \citet{gavriil}, which does display strong harmonic structure).  It therefore seems that the majority of the magnetars are neither 
nearly aligned nor nearly orthogonal.  However, one caveat needs to be added: the argument above,  while safe for, say, 
a radio pulsar, where the emission comes from high magnetospheric altitudes, is less secure for a magnetar.  
As described in \citet{opk01}, if the magnetar radiation is produced at the stellar surface itself, the combination of beaming and 
relativistic light bending can result in even a single hot spot profile giving rise to a pulse with multiple peaks, 
making statements concerning the inclination angle very difficult to make.  
%Indeed, \citet{opk01} find that a sinlge hot spot is a better fit to the pulse profile data that two anti-podal hotspots.   
%What this means for our assumption that the magnetar pulsation axis coincides with a magnetic deformation axis in not at all clear.  
We regard the assumption of non-orthogonality between the spin and magnetic axes as the weakest link in our argument.  
 
Ours is not the first analysis of the precessional dynamics of magnetars. As noted above, \citet{melatos} attempted to interpret the
irregular spin-down of AXPs as free precession, but sustained observations did not confirm his model 
\citep{kaspi1,kaspi2,gavriil}.  More closely related to our work is that of 
\citet{wasser}, who applied the same techniques as us (albeit without accounting for the presence of a superfluid component) 
to work out conditions under which radio pulsars would not precess, and used these to made 
statements about the conditions under which precession is likely, and how the star's shape may change plastically over long timescales.  
As outlined above, the problem we address in this paper is different and perhaps simpler. 
The observations indicate that the magnetars do not precess, which we then translate into constraints on the stellar structure.

%%%%%%%%%%%%%%%%%%%%%%%%%%%%%%%%%%%%%%%%%%%%
\section{Basic model}
\label{sect:bm}

We will begin by assembling the most basic model that illustrates the key features of
our argument. We assume a two-component neutron star model, consisting of a ``crust'' and a pinned neutron superfluid.  
In particular, the ``crust'' 
component includes the actual elastic crust, the charged fluids in the core as well as the portion of the neutron 
fluid not locked into the 
pinned component  (vortex mutual friction would bring the non-pinned superfluid into corotation with the charged component 
\citep{als88,na06}). The two components are allowed to spin with different angular velocities $ \Omega^i_\rC $ and 
$ \Omega^i_\rs $. 
The former can be identified  with the observed spin frequency $ \Omega^i = \Omega \n^i_\Omega $ (a ``hat'' denotes a unit vector).   
The crust's moment of inertia tensor can then be written
\be
I_\rC^{ij} = I_\rC \delta^{ij} + \Delta I_\rC \nd^i \nd^j + \Delta I_\rB \nB^i \nB^j .
\ee
This includes a spherical piece $I_\rc $ and the quadrupolar deformations due to the elastic crust ($\Delta I_\rC$) 
and the interior magnetic field ($\Delta I_B$). The two unit vectors $\nd^i, \nB^i $ are assumed fixed in the crust frame, and
we denote as $\beta$ the angle between them. For the superfluid component we will assume a spherical
moment of inertia tensor 
\be
I^{ij}_\rs = I_\rs \delta^{ij}.
\ee
We will make the perfect pinning assumption, i.e. the pinned neutron vortices are physically
immobilised with respect to the crust which in turn implies a superfluid angular velocity $\Omega_\rs^i = \Omega_\rs \nS^i$ 
which is fixed in the crust frame.

It will be useful to parameterise the sizes of these contributions to the moment of inertia tensor.  
If $I_0 = I_\rC + I_\rs$ denotes the spherical moment of inertia of the whole star, we can define
\be
\label{eq:epsilons}
\ed = \frac{\Delta I_\rC}{I_0}, \qquad \eB = \frac{\Delta I_B}{I_0}, \qquad 
\eS = \frac{ I_\rs }{I_0} .
\ee 

The deformation $\ed$ is sourced by strains in the solid crust. The recent calculation by \citet{bryn1}
place upper limits on this of
\be
 (\ed)_{\rm max} \approx 10^{-5} \left(\frac{u_{\rm br}}{10^{-1}}\right) ,
\label{ecmax}
\ee
where we have parameterised the crustal breaking strain $u_{\rm br}$ in terms of the value found in the state-of-the-art calculations 
of \citet{horowitz}.

For the magnetic deformation a somewhat crude estimate (but consistent with more rigorous
calculations, e.g. \citet{akgun, anto, lj09}) is
\be
\eB \approx 10^{-6} \left ( \frac{H}{10^{15}~\mbox{G}} \right ) \left ( \frac{\bar{B}}{10^{15}~\mbox{G}} \right ) ,
\label{eB} 
\ee
where $\bar{B} $ is a volume average of the internal magnetic field and $ H \approx 10^{15}\, {\rm G} $ if the 
core sustains type II proton superconductivity, otherwise $ H = \bar{B} $. The deformation $\eB$ can be
either positive or negative %(i.e. oblate or prolate respectively) 
depending on the relative strength between the poloidal and toroidal magnetic field components \citep{mestel,cutler}.

Clearly, given the previous estimates, the  magnetic and elastic deformations are small numbers; the same is not necessarily true for
$\eS$. If pinning occurs only in the inner crust, only a few percent of the stellar superfluid contributes 
to $I_\rs$, giving $\eS \approx 10^{-2}$.  However, if vortex pinning is efficient in the neutron star core due to the  
interaction with the magnetic fluxtubes then $\eS$ could even be of order unity.  
Either way, $\eS$ is likely to be much greater than either of $\ed$ and $\eB$.

We will now easily find the non-precessional solutions. The total angular momentum of our star is given by the sum of the crustal and 
superfluid contributions 
\be
J^i = I_\rC^{ij}  \Omega_j + I_\rs^{ij} \Omega^\rs_j .
\ee
Let $\theta_\rC$ and $\theta_\rB$ denote the angles that the crustal and magnetic deformations make with the angular velocity vector, 
so that $\nd^i \n^\Omega_i = \cos\theta_\rC$ and  $\nB^i \n^\Omega_i = \cos\theta_\rB$.  The angular momentum can then be written 
\be
J^i = \Omega \left ( I_\rc \n_\Omega^i + \Delta  I_\rc \cos\theta_\rc \nd^i  + \Delta  I_\rB \cos\theta_\rB \nB^i
\right ) 
+ I_\rs \Omega_\rs \nS^i .
\ee 
The Euler equations of rigid body dynamics then give
\be
\dot{J}^i + \epsilon^{ijk} \Omega_j J_k = 0 ,
\label{euler}
\ee
where the time derivative is evaluated in the rotating crust frame.  For a non-precessional and therefore stationary 
solution we must have $\dot{J}^i = 0$ which implies that $J^i$ must be parallel to $\Omega^i$, i.e. there exists a 
scalar $\lambda$ such that $J^i = \lambda \Omega \n_\Omega^i$.  We therefore obtain the relationship
\be
\label{eq:4_unit_vectors}
(I_\rc-\lambda) \nOm^i + \Delta I_\rc \cos\theta_\rc \nd^i + \Delta I_\rB \cos\theta_\rB \nB^i
+ I_s \frac{\Omega_s}{\Omega} \nS^i = 0
\ee
between the four unit vectors $\nd, \nB, \nOm, \nS$.  

Equation (\ref{eq:4_unit_vectors}) is crucial--we can easily derive our two key results from it.  
Firstly, consider the case where there is no pinned superfluid. Setting $I_\rs =0$ we obtain 
\be
\label{eq:3_unit_vectors}
\frac{I_\rc -\lambda}{I_0} \nOm^i + \ed \cos\theta_\rc \nd^i + \eB \cos\theta_B \nB^i  = 0 .
\ee
This shows that the three unit vectors $\nd, \nB, \nOm,$ are coplanar.  We can therefore make the decomposition 
into a pair of basis vectors $\n_1^i, \n_3^i$ such that
\bear
\hat{n}_{\rm B}^i &=& \sin\theta_{\rm B} \n_1^i + \cos\theta_{\rm B} \n_3^i ,
\\\
\nd^i &=& \sin\theta_\rc \n_1^i + \cos\theta_\rc \n_3^i ,\\
\nOm^i  &=& \n_3^i .
\eear
Inserting into equation (\ref{eq:3_unit_vectors}) and projecting along $\n_1^i$ leads to
the non-precession condition
\be
\label{eq:zero_pinning}
\ed \sin 2\theta_\rc = -\eB \sin 2\theta_\rB .
\ee
Thus, in the case of zero pinning, non-precession implies 
 a simple  relation between the crustal and magnetic deformations. The same result was obtained by \citet{wasser}. 

Now consider the case where there is pinning. 
We define the angle $\nS^i \n^{\rm \Omega}_i = \cos\theta_\rs$. Taking the vector product of equation (\ref{eq:4_unit_vectors}) with $\nOm^i$ 
and squaring the result leads to
\begin{multline}
\left (2 \eS\frac{\Omega_\rs}{\Omega} \sin \theta_s \right )^2 = (\ed \sin2\theta_\rc)^2 
+  (\eB \sin2\theta_\rB)^2
\\
+  8 \ed \eB \cos\theta_\rc \cos\theta_\rB (\cos\beta - \cos\theta_\rc \cos\theta_\rB) .
\label{eq:pinning}
\end{multline}
This is a rather cumbersome relation  between the angles that the superfluid, the crustal deformation, and the magnetic axis make 
with the rotation axis.  However, from this we see one important thing: the star rotates about an axis very close to the superfluid  
pinning axis, with the angle of misalignment being of order of the larger of $\Delta I_\rc/I_\rs$ and $\Delta I_\rB/I_\rs$.  
In the notation of (\ref{eq:epsilons}), we therefore have
\be
\label{eq:theta_s}
\theta_\rs \sim \max{ \left(\frac{\ed}{\eS}, \frac{\eB}{\eS}  \right) } \ll 1 .
\ee
Note that in the pinning scenario, the magnetar non-precession does not place any constraint on the crustal deformation.  
Also, it is possible for the star to spin about the crustal principal axis (as defined by eqn.\  (\ref{eq:zero_pinning})) 
and have a pinned superfluid, provided the pinning axis lies \emph{exactly} along the spin axis ($\theta_s = 0$).  
This would correspond to both our scenarios being realised. However, there is no apriori reason to expect the symmetry axis of the 
pinning sites to coincide with the spin axis, so this is a clearly special case.

It is also interesting to note the well known result  that the non-precessing state is the one that mimimises the kinetic energy at 
fixed $J$  \citep{ll76,shaham77}. Therefore,  an isolated star, dissipating kinetic energy, would evolve to 
such a state on the dissipation timescale.

Finally, note that this last point has repercussions for the so-called Mestel-Jones spin-flip mechanism 
\citep{mestel, jones75, cutler}, in which a star with a dominantly prolate magnetic deformation of its inertia tensor 
(i.e. $\eB <0$) is driven to a configuration with the magnetic axis orthogonal to the spin axis. 
%(i.e. $\theta_{\rm B} \rightarrow \pi/2$).  
Clearly, the presence of a pinned superfluid 
completely suppresses this, assuming that the magnetic and pinning axes remain fixed with respect to each other 
(as in our model).   According to the above discussion,  the minimum energy state for a star containing a pinned superfluid 
has $\theta_s \approx 0$, i.e. the pinning axis must remain close to the spin axis, preventing the migration of the magnetic axis to 
an orthogonal location.

%%%%%%%%%%%%%%%%%%%%%%%%%%%%%%%%%%%%%%%%%%%
\section{Extensions to the basic model}
\label{sect:ettbm}

The previous model can be extended in terms of realism by incorporating
additional physical effects. These include (i) rotational ``bulges'' in the moments of inertia, and (ii) the action of a non-dissipative torque due to the
external dipole field (the so-called ``anomalous'' torque \citep{goldreich}).

With the rotational deformations included the previous moment of inertia tensors become
\bear
&& I_\rC^{ij} = I_\rC \delta^{ij} + \Delta I_{\rm \Omega} \nOm^i \nOm^j 
+ \Delta I_\rC \nd^i \nd^j + \Delta I_\rB \nB^i \nB^j ,
\\
&& I^{ij}_\rs = I_\rs \delta^{ij} + \Delta I_\rs \nS^i \nS^j .
\eear
For a slowly rotating object like a magnetar we would expect $\Delta I_{\rm \Omega} \ll I_\rc$ and
$\Delta I_\rs \ll I_\rs$. 
The total angular momentum is
\begin{multline}
\label{eq:J_total}
J^i = \Omega \left \{ \left ( I_\rc +  \Delta I_{\rm \Omega} \right ) \n_\Omega^i + \Delta  I_\rc \cos\theta_\rc \nd^i  
\right.
\\
\left. + \Delta  I_\rB \cos\theta_\rB \nB^i \right \}
+ \left ( I_\rs + \Delta I_\rs \right ) \Omega_\rs \nS^i .
\end{multline}
This expression shows that the centrifugal deformations are not expected to play any significant
role since they are simply ``absorbed'' in the pre-existing spherical pieces. 

The second modification to the basic model, the anomalous torque $N^i_{\rm an}$, will appear as a source term in the 
Euler equation (\ref{euler}).  If the exterior space is assumed perfect vacuum and the magnetic 
field  dipolar this torque can be written as
\be
N^i_{\rm an} = I_0 \Omega \eA ( \nOm^l \m_l ) \epsilon^{ijk} \Omega_j \m_k  ,
\ee
where $m^i$ is the \emph{external} magnetic dipole moment and $\eA$ is an effective ``deformation''
\be
\eA = \frac{m^2}{I_0 R c^2} ,
\ee
and $R$ is the stellar radius. Expressed in terms of the field strength $B_d$ on the magnetic pole
we obtain the estimate 
\be
\eA \approx 10^{-7} \left ( \frac{B_d}{10^{15}\,{\rm G}} \right )^2 .
\ee
Clearly, the impact of the anomalous torque on the stellar dynamics is 
important only for magnetar-strength fields. This was the central idea in the so-called radiative 
precession model advocated by \citet{melatos}.

With the inclusion of the anomalous torque the equation of motion (\ref{euler}) takes the form
\be
\dot{J}^i + \epsilon^{ijk} \Omega_j \left [  J_k -\eA I_0 \Omega  ( \nOm^l \m_l ) \m_k \right  ]   = 0 .
\ee
Precession will not occur provided
\be
J^i = \lambda \Omega \nOm^i + \eA I_0 \Omega  ( \nOm^j \m_j ) \m^i .
\label{noprecess}
\ee
It is natural to assume that the interior and exterior magnetic fields share the same symmetry axis
(see discussion in the next section). Setting $\m^i=\nB^i$ in (\ref{noprecess})  
we recover the previous  results (\ref{eq:zero_pinning})-(\ref{eq:theta_s}) with
the recalibrated deformations $ \epsilon_\rB \to \eB -\eA $ 
and $ \epsilon_\rs \to  \eS + \Delta I_\rs/I_0$.

%%%%%%%%%%%%%%%%%%%%%%%%%%%%%%%%%%%%%%%%%%%%
\section{Discussion}
\label{sect:discussion}

First consider equation (\ref{eq:zero_pinning}), which applies in the limit of no pinning. 
The magnetic ellipticity $\eB$ is at least as large as would be estimated by substituting the 
\emph{external} magnetic field strength into equation  (\ref{eB}), i.e. for a typical magnetar $\eB \sim 10^{-6}$
for $\bar{B} = 10^{15}\, \mbox{G}$.
In reality this number could be significantly larger if we take at face value the most
recent calculations of stable magnetic equilibria in neutron star models which suggest that the interior field $\bar{B}$
could be much stronger than the exterior field $B_d$ \citep{braith09}. 

That the magnetars do not seem to be either aligned or orthogonal rotators means that the angle $\theta_B$ is neither 
close to zero nor close to $\pi/2$, and so the trigonometric factor on the right hand side of (\ref{eq:zero_pinning}) is 
of order unity.  
It then follows that the crustal deformation must be of the order of the magnetic one: $\ed \sim 
\eB \gtrsim 10^{-6}$, or possibly even larger, if $\theta_{\rm c}$ is small.  The value $\ed \sim 10^{-6}$
is rather significant-- it is close to the maximum allowed deformation predicted by detailed modelling of the crust,
i.e. equation~(\ref{ecmax}). 

It follows that, in the absence of superfluid pinning, magnetar crusts are maximally, 
or close to maximally, strained. This idea sits well with the standard model of magnetar activity
\citep{TD} where the energy budget of the flaring activity is supplied by the evolving magnetic field.  
The crust itself acts as a gate, with crustal cracking acting as a trigger for the magnetic reconnection events.  
The maximally strained crust that applies in this scenario makes such events perfectly natural.

One might worry that one could evade our conclusion of a strongly strained crust by relaxing the assumption that the internal 
and external magnetic fields are aligned (i.e. $\hat{m}^i \neq \hat{n}^i_B$).  
Indeed, if the \emph{internal} magnetic axis lies close to the rotation axis 
(making the angle $\theta_\rB$  small), equations (\ref{eq:zero_pinning}) allow for a $\ed$
well below the maximum.  
However, a misalignment between the internal 
and external magnetic fields will generate tangential magnetic stresses, which can only be balanced by elastic ones.  
If both the internal and external fields are of strength $\sim B$, and if the field transitions from one geometry to 
the other over a shell of thickness $\Delta R < R$, the tangential magnetic stresses will be the order of $B^2/\Delta R$.  
This stress is greater than the  $B^2/R$ that the crust needs to balance in our original  scenario with aligned internal and 
external fields. Clearly, relaxing this assumption does not allow one to avoid our conclusion of a highly strained crust.
Additional evidence supporting the alignment between the interior and exterior magnetic fields comes from simulations 
of stable magnetic equilibria \citep{braith09}.

Now consider the case with superfluid pinning (equations (\ref{eq:pinning}) and (\ref{eq:theta_s})). 
In this case no constraint is placed on the crustal deformation.  In fact, $\ed$ could be set to zero, 
with no qualitative change in the non-precessional solution.
The only constraint on stellar parameters that the pinning scenario would imply is that 
the pinning force itself is strong enough to sustain the small misalignment between the crustal and superfluid angular velocity 
vectors given by equation (\ref{eq:theta_s}).  As discussed by \citet{link_cutler} 
the relative velocity between the superfluid and the crust generated by this misalignment produces a Magnus 
force on the pinned vortices. Making a simple estimate, the relative velocity generated by the misalignment is 
of order $\Delta v \sim \Omega R \theta_\rs$.
Parameterising with magnetars in mind gives
\be
 \label{eq:v_critical_param}
  \Delta v \sim 60 {\, \rm cm/s} 
  \left(\frac{\Omega}{0.63 \rm Hz}\right)
  \left(\frac{\Delta I/I_0}{10^{-6}}\right)
  \left(\frac{10^{-2}}{\eS}\right) .
\ee
This is to be compared with the critical velocity for unpinning.  
The estimates of this have been collected in \citet{dij09} who finds a lower unpinning threshold 
$\sim 10^4\,\mbox{cm}/\mbox{s}\gg \Delta v$.
Thus, the finite strength of vortex pinning is no 
obstacle to building our non-precessing magnetar models. Furthermore, the weak $\Delta v$ also ensures that 
the non-precession configuration is immune to the superfluid instability discussed in \citet{precpaper}.

There is one special case that deserves some comment: when the crustal and magnetic axes are aligned, so that $\nd^i = \nB^i$.  
In this case the star is biaxial rather than triaxial, and precession would correspond to the magnetic axis rotating in 
a cone of half-angle $\approx \theta_B$  about the fixed vector $J^i$, with a rotation about $\nB^i$ at the precession 
frequency superimposed.  As is well known, such a precession would have no signature in the pulsations \citep{ja01}.  
Such a seemingly non-precessing star could have no pinned superfluid and any amount of crustal strain (even zero--in this case the star is
still biaxial), and so would not fit into 
the scheme described above.  How can we be sure that magnetars do not fall into this special case?  Firstly, the absence of 
precessional modulation in the pulsations would require the hot-spot itself to be exactly axisymmetric about the magnetic axis.  
Secondly, it would be surprising if the crustal strains were symmetric about the magnetic axes, as the crustal strain field is 
likely to have retained some memory of a more oblate shape inherited from earlier in its  life, before it was spun-down to the 
long periods of the magnetars.  Thirdly,  such a large-amplitude precession would be expected to damp due to internal 
dissipative processes \citep{ja01}.  We therefore feel it is very unlikely that this special case is common in the magnetar population.

Having arrived at two different non-precession possibilities, described by equations (\ref{eq:zero_pinning}) and 
(\ref{eq:pinning}), it is natural to ask the question of how one might hope to distinguish between them.   
As already noted, the no pinning case, with its maximally strained crust, sits naturally with the magnetar flare model, 
as quake events in the crust are required to trigger magnetic reconnection events.  However, this is hardly 
decisive--diffusion of the magnetic field may, in either scenario,  play a key role in triggering the bursts.

Nevertheless, one intriguing possibility does suggest itself. In the scenario with no superfluid pinning the 
crustal strain predicted just happens to lie at the upper end of what is believed to be possible on the basis of detailed 
crust modelling. Suppose this is no coincidence.  It may be the case that there exist even more strongly magnetised stars; 
their magnetic deformations 
would necessarily be larger than their crustal ones.  By eqn.~(\ref{eq:zero_pinning}) it follows that such stars, provided they are oblate
($\epsilon_\rB > 0$), would be nearly aligned rotators.  
There would be a selection effect against seeing such magnetars.   
So, observation of a tail of strongly magnetised 
but nearly aligned rotators would argue in favour of there being no superfluid pinning  in magnetars.   
Similarly, if the magnetic deformation were prolate ($\epsilon_\rB < 0$), strongly magnetised stars would be orthogonal rotators, 
although as noted above, the current observations do not suggest this to be the case.  Note, however, regardless of the sign 
of $\epsilon_\rB$,   the absence of such aligned/orthogonal populations would not necessarily imply that superfluid pinning 
occurs in magnetars;  Nature may simply  not provide neutron stars with magnetic field strengths higher than those currently observed.

The presence of a pinned superfluid is also related to the timing profile of magnetars.
Indeed, several AXPs are known to have undergone one or more glitches with a fractional $\Omega$-jump comparable to the 
Vela-type glitches in radio pulsars (for a review see \citet{kaspi}). 
The occurrence of large glitches by itself suggests the presence of a pinned superfluid reservoir somewhere in the stellar 
interior, capable of transfering angular momentum to the crust if some mechanism could cause vortex unpinning.  
So far no glitches have been detected in SGRs. This should not come as a surprise
given that these objects, when in quiescence, are actually fainter than the AXPs as well as ``louder'' in terms
of timing noise. As a result, the phase-coherent timing of SGRs required for the identification of glitches becomes
a highly problematic procedure. %\footnote{We thank V. Kaspi for pointing this out to us}. 
We can, nevertheless, speculate on physical grounds as to why SGRs may not glitch.     
One possibility could be the absence of vortex pinning in the core, \emph{if}  the core is the only region where pinning could in 
principle occur. The interior magnetic field in SGRs could be sufficiently
strong as to exceed the critical field $H_{\rm c2} \approx 10^{16}\,{\rm G}$ above which proton superconductivity
is suppressed \citep{baym}. The destruction of superconductivity would obviously imply the absence of magnetic fluxtubes on which
the vortices could pin. If this scenario were to be true then the known SGRs would obey the non-precession condition
(\ref{eq:zero_pinning}) and would have maximally strained crusts.

It is interesting to consider the impact that a hypothetical discovery of magnetar precession would have
on our principal conclusions.  As we discussed in the introduction, current observations do not exclude the possibility that 
magnetars could be undergoing small-amplitude precession, masked by their timing noise. This precession could be either fast
or slow (i.e. with a period, respectively, comparable to or much longer than the spin period) 
depending on whether there is a pinned superfluid. A detection of the former type of precession
would amount to clear evidence of the existence of a pinned superfluid component within the star. 
Alternatively, detection of slow precession would mean that (i) there is no pinned
superfluid, and (ii) that the star's rigid body dynamics deviates only slightly from the non-precession condition (\ref{eq:zero_pinning}),
if the precession amplitude is sufficiently small. Thus, even in this case we would be able to make the same predictions 
for the magnetar structure as before.

To sum up, we have argued that the high magnetic strains that must necessarily exist in magnetars  
have interesting implications for their structure, telling us something about the level of crustal strain or the superfluid 
state of matter in their interior.  
Our modelling assumes a lack of significant free precession in the magnetars, and that their magnetic axes, 
as traced by the pulsation axes, are neither nearly aligned nor nearly orthogonal to their spin axes.  
We therefore conclude by pointing out the importance of better observational data to either confirm or reject these assumptions.  
In particular, our work shows that there would be a great deal of interest in attempting to construct detailed models 
of the geometry of magnetar emission, despite the technical difficulties involved.
There would also be interest in searching for an ultra-highly magnetised but nearly aligned or orthogonal tail to the 
magnetar population. Clearly, more detailed observations of magnetar timing and pulse geometry could give us a unique insight 
into neutron star interiors.

%%%%%%%%%%%%%%%%%%%%%%%%%%%%%%%%%%%%%%%%%%%%%%%%%%%%%%%%

\section*{Acknowledgments}
KG is supported by the German Science Foundation (DFG) via SFB/TR7.
DIJ is supported by STFC via grant number PP/E001025/1.  The authors
also acknowledge support from COMPSTAR (an ESF Research Networking
Programme) and thank Vicky Kaspi, Alessandro Patruno and Anna Watts for
valuable feedback on a first draft of this paper.

%%%%%%%%%%%%%%%%%%%%%%%%%%%%%%%%%%%%%%%%%%%%%%%%%%%%%%%%%%%%%

%%%%%%%%%%%%%%%%%%%%%%%%%%%%%%%%%%%%%%%%%%%%%%%%%%%%%%%%%%%%%%%%%%%%%

\end{document}